\RequirePackage{fix-cm}
\documentclass[smallextended]{svjour3}       
\smartqed  
\usepackage{graphicx}
\usepackage{cite}
\usepackage{amsmath,amssymb,amsfonts}
\usepackage{algorithmic}
\usepackage{graphicx}
\usepackage{textcomp}
\usepackage{xcolor}
\usepackage{booktabs}
\usepackage{multirow}
\usepackage{natbib}
\usepackage{url}
\usepackage{tcolorbox}
\begin{document}

\title{COBOL-Coder: Domain-Adapted Large Language Models for COBOL Code Generation and Translation
}

\titlerunning{Domain-Adapted Large Language Models for COBOL Code Generation}        

\author{Anh T. V. Dau         \and
        Shin Hwei Tan  \and
        Jinqiu Yang \and
        Nghi D. Q. Bui \and
        Anh Tuan Nguyen
}


\institute{Anh T. V. Dau \at
              Concordia University, Montreal, Canada \\
              \email{thivananh.dau@mail.concordia.ca}           
           \and
           Shin Hwei Tan \at
              Concordia University, Montreal, Canada \\
              \email{shinhwei.tan@concordia.ca} 
            \and
            Jinqiu Yang \at
              Concordia University, Montreal, Canada \\
              \email{jinqiu.yang@concordia.ca} 
            \and
            Nghi D. Q. Bui \at
              FPT Software AI Center, Vietnam \\
              \email{nghibdq@fpt.com} 
            \and
            Anh Tuan Nguyen \at
              FPT Software AI Center, Vietnam \\
              \email{anhnt446@fpt.com} 
}

\date{Received: date / Accepted: date}

\maketitle

\newcommand{\modelname}{\textsc{COBOL-Coder }}
\newcommand{\modelnamenospace}{\textsc{COBOL-Coder}}
\newcommand{\benchmarkname}{\textsc{COBOL-JavaTrans }}
\newcommand{\benchmarknamenospace}{\textsc{COBOL-JavaTrans}}
\newcommand{\todo}[1]{\textcolor{red}{{\it [TODO: #1]}}}
\newcommand{\shin}[1]{\textcolor{red}{{\it [SHIN: #1]}}}
\newcommand{\jinqiu}[1]{\textcolor{cyan}{{\it [Jinqiu: #1]}}}
\newcommand{\fix}[1]{\textcolor{blue}{{\it [Anh: #1]}}}

\begin{abstract}
COBOL remains a critical language for mainframe systems, yet existing large language models (LLMs) struggle to generate and translate COBOL code correctly. This paper reports our experience in developing and evaluating domain-adapted LLMs for COBOL and mainframe software engineering. We introduce (1) an automated data curation pipeline that combines compiler-guided validation with multi-stage similarity-based filtering to construct high-quality COBOL training data, 
and (2) \modelnamenospace, a COBOL-specialized LLM fine-tuned on the curated COBOL domain data. 
 We evaluate \modelname to two tasks: code generation (on COBOLEval and COBOLCodeBench) and code translation (on \benchmarknamenospace, our proposed benchmark for bidirectional COBOL–Java translation). In our experiment, \modelname achieves up to a 73.95\% compilation success rate and 49.33 Pass@1 on COBOLEval, compared to 41.8\% and 16.4 for GPT-4o, while most open-source baselines (e.g., CodeGemma, CodeLlama, StarCoder2)
 fail to produce compilable programs. For Java-to-COBOL translation, \modelname reaches 34.93 Pass@1, whereas general-purpose LLMs achieve near-zero scores. 
To assess the usability of LLM-generated code in real-world settings, we conduct a survey with experienced COBOL developers. Participants consistently report that \modelname exhibits stronger COBOL awareness, has more reliable program structure, and is better aligned with enterprise practices than general-purpose LLMs.
\end{abstract}

\section{Introduction}
Large language models (LLMs) have recently become effective tools for software engineering tasks such as code generation, translation, and maintenance \citep{chen2021codex,team2024qwen2,gpt4,guo2025deepseek}. Trained on large-scale code corpora, modern LLMs achieve strong performance on widely used programming languages, often approaching or surpassing human-level results on standard benchmarks. However, their effectiveness drops significantly when applied to legacy and low-resource programming languages. COBOL, which continues to underpin mission-critical systems in banking, insurance, and government \citep{taulli_cobol_nodate}, exemplifies this gap. Despite its importance, existing LLMs perform poorly on COBOL-related tasks. On COBOLEval \citep{noauthor_bloop_nodate}, an adaptation of HumanEval for COBOL, the top-performing model achieves only 10.37 Pass@1. Similar limitations are observed in other low-resource languages such as Fortran, where even advanced LLM-assisted migration remains challenging due to unresolved dependencies and inconsistencies \citep{joel2024survey}. These results highlight a fundamental limitation of current LLMs in handling legacy programming languages.

Producing LLMs for legacy programming languages, especially COBOL, remains challenging due to several limitations. First, such languages are inherently low-resource: most production code resides in enterprise systems and is rarely publicly available, leading to limited training data. As a result, LLMs 
trained primarily on modern languages struggle to learn legacy-specific syntax, programming idioms, and domain conventions \citep{dau2024xmainframe}. Second, although initial benchmarks such as COBOLEval \citep{noauthor_bloop_nodate} have begun to support evaluation for legacy language code generation, the existing benchmark landscape remains narrow in both task diversity and practical coverage. In particular, current datasets offer limited support for broader software engineering tasks such as code translation and program modernization.
Third, legacy languages often differ substantially from modern programming paradigms in terms of program structures, formatting requirements, and data layouts, which makes code generation particularly challenging. Moreover, to date, relatively limited research effort has been devoted to systematically investigating LLMs for legacy programming languages, which are inherently low-resource and lack large-scale training corpora. These challenges motivate the need for specialized datasets, evaluation frameworks, and modeling approaches tailored to legacy software systems.

In this work, we investigate whether COBOL-specific adaptation improves LLMs' performance on COBOL tasks.
We first introduce a data augmentation pipeline that generates high-quality COBOL training data through compiler-guided validation and multi-stage similarity-based filtering, enabling the construction of reliable instruction data for both code generation and translation tasks. 
Building on these resources, we develop \modelnamenospace, a COBOL-specialized LLM fine-tuned on the curated dataset. We further present \benchmarknamenospace, the first benchmark for bidirectional COBOL–Java translation. Finally, we complement our evaluation with a user study involving experienced COBOL developers to assess the practical utility of LLM-generated code in real-world scenarios.

To evaluate the effectiveness of \modelnamenospace, we conduct experiments on both code generation and code translation tasks. For code generation, we evaluate them on existing COBOL benchmarks and show that
many widely used code LLMs fail to generate executable programs, while our
LLMs achieve substantially higher compilation success and correctness. For code translation, we introduce \benchmarknamenospace, a benchmark for bidirectional COBOL–Java translation derived from HumanEval \citep{chen2021codex}. Our results show that general-purpose LLMs achieve near-zero performance on COBOL tasks, while our LLMs attain non-trivial results. In addition to automated evaluation, we assess the practical utility of LLM-generated code through a user study with experienced COBOL developers. Participants evaluate model outputs for both generation and translation tasks, providing qualitative and task-level feedback on correctness, readability, and adherence to established COBOL programming practices.

Our evaluation aims to address the research questions below:

\noindent\textbf{RQ1: How well do \modelname and existing LLMs generate compilable and correct COBOL code?} We evaluate COBOL code generation on COBOLEval \citep{noauthor_bloop_nodate} and COBOLCodeBench \citep{cobolcodebench} using compilation success rate (CSR) and functional correctness (Pass@1). On COBOLEval, most open-source baselines, including CodeGemma \citep{team2024codegemma}, CodeLlama \citep{roziere2023code}, StarCoder2 \citep{lozhkov2024starcoder}, and DeepSeek-R1-Distill-Qwen \citep{guo2025deepseek}, achieve 0\% CSR and 0\% Pass@1 on COBOLEval and COBOLCodeBench, while the existing best-performing GPT-4o \citep{gpt4o} reaches at most 41.8\% CSR and 16.4 Pass@1. In contrast, \modelnamenospace-7B achieves 73.80\% CSR and 44.70 Pass@1, and \modelnamenospace-14B further improves to 73.95\% CSR and 49.33 Pass@1. Notably, \modelnamenospace-14B is the only model with non-trivial performance on COBOLCodeBench (26.09\% CSR, 4.35 Pass@1).

\noindent\textbf{RQ2: How effective is \modelname at translating between COBOL and Java in both directions on \benchmarknamenospace?} On COBOL-to-Java translation, \modelname outperforms all open-source LLMs, achieving 97.9\% CSR and up to 83.91 Pass@1, approaching much larger general-purpose LLMs. The Java-to-COBOL translation task is substantially harder: most LLMs fail completely, while \modelname achieves 63.64\% CSR / 27.27 Pass@1 on the 7B version and 72.03\% CSR / 34.93 Pass@1 on the 14B version, clearly outperforming all baselines.

\noindent\textbf{RQ3: How do experienced COBOL developers evaluate the quality and practical usefulness of LLM-generated code in realistic development scenarios?}
We conduct a study with experienced COBOL developers to evaluate the practical usefulness of LLM-generated code in realistic development scenarios. Participants assessed outputs from three LLMs, including \modelnamenospace, and widely used general-purpose LLMs—GPT-4 and GPT-4o—across tasks covering COBOL code generation and bidirectional translation between COBOL and Java, ranking each solution based on functional correctness, readability, and adherence to conventional COBOL coding practices. Overall, \modelname is ranked first across all Java-to-COBOL translation tasks and first or tied for first in most COBOL code generation tasks. Qualitative feedback indicates that \modelname is perceived as more production-oriented and COBOL-aware, particularly in terms of program structure and alignment with enterprise development patterns. 

In summary, our work makes the following contributions:
\begin{itemize}
\item We propose an automated data augmentation pipeline, which produces a large-scale COBOL-specific instruction-tuning corpus for the fine-tuning process.

\item We introduce \modelnamenospace, an LLM specialized for COBOL-related tasks and conduct evaluations on both COBOL code generation and translation. Our results show that \modelname outperforms all current state-of-the-art open-source LLMs and GPT variants, surpassing them in terms of compilation success and functional correctness. 

\item We construct \benchmarknamenospace, the first benchmark for COBOL–Java code translation, targeting practical legacy system modernization scenarios where COBOL programs are migrated to modern languages.



\item We conduct a survey-based evaluation with experienced COBOL developers, providing qualitative insights into the usability, readability, and correctness of LLM-generated legacy code. 

\end{itemize}


\section{Methodology}
In this section, we describe the data augmentation pipeline used to construct the training data in Section \ref{sec:data}, followed by the instruction-tuning procedure presented in Section \ref{sec:instruction}.

\subsection{Curation of COBOL-specific Training Data}
\label{sec:data}

We construct the training data for \modelname using three complementary sources: (\emph{Source 1}) real-world COBOL programs collected from GitHub (Section \ref{sec:source_code}), (\emph{Source 2}) synthetic COBOL programs generated via cross-language translation (Section \ref{sec:synthetic_data}), and (\emph{Source 3}) COBOL and mainframe knowledge sources (Section \ref{sec:document}),
as illustrated in Figure \ref{fig:code_pipeline}.




\subsubsection{Source 1: Public COBOL Code from Github}
\label{sec:source_code}

\begin{figure*}[ht]
\centering
\includegraphics[width=\textwidth]{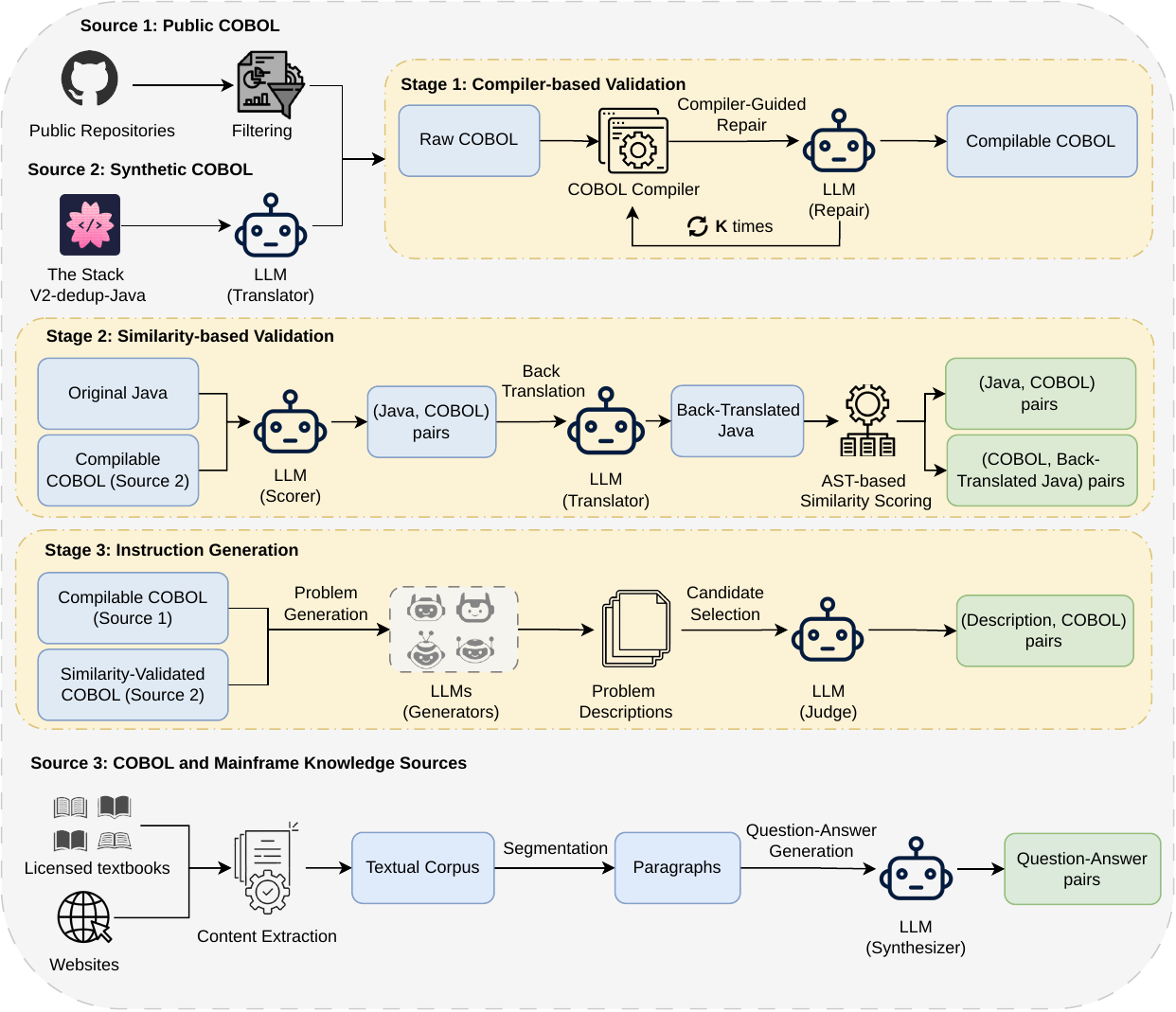}
\caption{Overview of the automated data augmentation pipeline.}
\label{fig:code_pipeline}
\end{figure*}

We mine real-world COBOL programs from public GitHub repositories using the GitHub Search API. Rather than collecting entire repositories, we adopt a file-level retrieval strategy by querying COBOL-related keywords (e.g., \textit{COBOL, COBOL-85, COBOL-2002}) in combination with standard file extensions (e.g., \textit{.cbl, .cob, .cobol, .cbx}). 
Archived and forked repositories are excluded as these repositories are either not actively maintained or duplicated repositories.

After data collection, we use a  filter to remove low-quality files. First, we discard files containing fewer than 20 lines of code. Such files frequently correspond to incomplete examples, configuration stubs, copybooks with minimal content, or placeholder files that contribute limited semantic value. Additionally, files exceeding 50,000 lines are excluded, as they are often the result of concatenated binaries or code conversion artifacts that deviate substantially from standard COBOL development practices.
We further eliminate non-text files and files with an unusually high proportion of non-alphanumeric characters, which typically indicate corrupted content or encoding issues. To mitigate redundancy and reduce the risk of memorization during model training, we incorporate a dedicated deduplication stage in the pipeline. We employ MinHash-based fingerprinting combined with Locality Sensitive Hashing (LSH) to identify near-duplicate COBOL files. Files with high content similarity are clustered together, and only a single representative file is retained from each cluster. This step removes duplicated source files that arise from code reuse across repositories while preserving diversity in coding patterns and program logic. 

After completing the above preprocessing steps, the corpus consists of 40,829 unique COBOL source files. These cleaned programs form the input to the first stage of our data construction pipeline in Figure~\ref{fig:code_pipeline}.



\noindent\textbf{Stage 1: Compiler-based Validation.}
A key challenge in constructing large-scale COBOL training corpora is that a substantial portion of real-world code does not compile out of the box. Public repositories frequently contain incomplete programs, outdated dialects, missing copybooks, inconsistent data definitions, or syntactic deviations across compiler implementations \citep{litecky1974study,ali2023x,lei2025enhancing}
. Naively discarding such files would significantly reduce data volume and bias the dataset toward overly clean or trivial examples. To address this issue, we implement a compiler-based validation process with a self-debugging mechanism, using compiler feedback to repair defective COBOL programs into syntactically correct training samples.

In this stage, each cleaned COBOL file obtained from the preprocessing step is first compiled using a standard COBOL compiler (GnuCOBOL~\footnote{https://gnucobol.sourceforge.io/}). If compilation errors occur, the compiler produces diagnostic messages in the form of a compilation log. The original source code and the corresponding compiler log are jointly provided to an LLM (Repair), instantiated using GPT-4o \citep{gpt4o}, which is instructed to act as an experienced COBOL developer to fix uncompilable programs. The LLM is tasked with (i) interpreting the compiler diagnostics, (ii) reasoning about the underlying causes of the reported errors, and (iii) generating a revised version of the program that resolves all compilation issues while preserving the original program intent. The corrected program is then recompiled to verify its validity. If compilation errors persist, the updated error messages are fed back to the LLM in a subsequent iteration, forming a compiler-in-the-loop self-debugging cycle. This process continues until either the program compiles successfully or a predefined maximum number of iterations $K=3$ is reached. Only programs that compile without errors are retained for downstream use. We provide all prompts used for LLM-based generation, translation, and evaluation in Appendix~\ref{appendix:prompt} to ensure reproducibility.

After this stage, we obtain 31,492 compilable COBOL programs, comprising approximately 38.4 million tokens in total, which form the core code corpus used for subsequent instruction construction and fine-tuning.
As these programs originate from real-world repositories, they are directly used for instruction generation without requiring additional semantic validation.

\noindent\textbf{Stage 3: Instruction Generation.}
For real-world COBOL programs, the compilable programs obtained after compiler-based validation are directly used as inputs to this stage. 
While Stage 1 ensures syntactic correctness, this stage focuses on generating high-quality problem descriptions that correspond to the given COBOL programs. 
Prior studies have shown that LLMs are effective at natural language generation and semantic abstraction, making them well-suited for synthesizing task specifications from source code \citep{yu2023large,wei2023magicoder,dau2024xmainframe}. Therefore, we leverage LLMs to transform validated COBOL programs into well-structured coding problems.

Given a validated COBOL program, we employ multiple LLMs to independently generate candidate problem descriptions. Each candidate includes a natural-language task specification, input/output requirements, and functional constraints inferred from the program behavior. In our implementation, we use GPT-4 \citep{gpt4}, GPT-4o-mini \citep{gpt4omini}, GPT-oss-120B \citep{gptoss}, and CodeLlama-70B \citep{roziere2023code} as generators to produce diverse candidate instructions for each program.

Next, we perform instruction validation and selection using an LLM as a judge. In this setting, the judge model does not generate new content but instead evaluates and ranks multiple candidate instructions corresponding to the same COBOL program. All candidate descriptions are provided to GPT-4o \cite{gpt4o}, which evaluates correctness, completeness, and alignment with the program logic to select the most faithful and informative description.
Each sample is then packaged into a standardized instruction format consisting of (i) a problem description specifying the task requirements, (ii) an input/output description detailing data formats and assumptions, (iii) the solution represented by the validated COBOL program, and (iv) a natural-language explanation summarizing the program logic. Through this process, we obtain high-quality instruction–solution pairs that support COBOL code generation during fine-tuning.

\subsubsection{Source 2: Synthetic COBOL Code via Code Translation}
\label{sec:synthetic_data}
Although COBOL source code is available in public repositories, its volume is orders of magnitude smaller than that of modern programming languages such as Java and Python. General-purpose code LLMs are typically trained on extremely large corpora: StarCoder2 \citep{lozhkov2024starcoder} is trained on approximately 3.3–4.3 trillion tokens of source code; the DeepSeek-Coder series \citep{guo2024deepseek} has approximately 2 trillion tokens sourced from 87 programming languages, and Magicoder \citep{wei2023magicoder} leverages more than 75K synthetic instruction examples in addition to large-scale pretraining. By comparison, the limited amount of publicly available COBOL code is insufficient for effectively fine-tuning LLMs toward COBOL-specific generation. This data scarcity motivates the need for synthetic data augmentation.

To mitigate this limitation, we generate synthetic COBOL data via cross-language translation, exploiting the semantic overlap between COBOL and modern languages. Translating programs from a high-resource language into COBOL enables the transfer of general programming patterns into the low-resource COBOL domain \citep{sontakke2023knowledge,gandhi2024translation}, substantially increasing both data volume and task diversity.
We use the Stack-v2-dedup-Java dataset\footnote{https://huggingface.co/datasets/bigcode/the-stack-v2-dedup} as the seed corpus, which is a filtered subset of The Stack v2 \citep{lozhkov2024starcoder} containing permissively licensed Java code that has undergone data cleaning and decontamination. 
Each Java program is translated into COBOL using an LLM (Translator), instantiated with GPT-4o, producing an initial COBOL version intended to preserve the original program functionality. 
However, due to structural and semantic differences between the two languages, these first-pass translations often contain syntactic errors and semantic inconsistencies \citep{gandhi2024translation,kumar2024automated,froimovich2025quality}.
To ensure syntactic correctness, the translated COBOL programs are first passed through the same compiler-based validation pipeline described in Section~\ref{sec:source_code}. Programs that fail to compile are iteratively repaired using an LLM (Repair) guided by compiler diagnostics until they become syntactically valid or reach the maximum number of iterations. After this stage, we obtain 279,536 syntactically valid Java–COBOL pairs, which serve as inputs to the subsequent semantic validation stage.


\noindent\textbf{Stage 2: Similarity-based  Validation.}
While Stage 1 ensures syntactic correctness, it does not guarantee that the translated COBOL programs preserve the behavior of the original Java programs. To address this limitation, we introduce a similarity-based validation stage that filters translation pairs based on cross-language consistency.
As shown in Figure~\ref{fig:code_pipeline}, this stage takes as input the original Java programs and the corresponding compilable COBOL programs obtained after Stage 1.
For each Java–COBOL pair, we perform two similarity checks. First, we apply LLM-based pair scoring to evaluate whether the COBOL program captures the functionality and logic of the original Java program. Second, we perform a back-translation procedure, where the COBOL program is translated back into Java and compared with the original program using AST-based similarity scoring. 

\noindent\textbf{LLM-based Pair Scoring:} 
Given a Java–COBOL pair, we apply a pair scoring step using an LLM (Scorer), which evaluates the similarity between the two programs. 
The model assesses whether the translated COBOL program correctly captures the functionality, control flow, and data transformations of the original Java code. The LLM is prompted to assign a similarity score in the range $[0,1]$ along with a brief explanation. Only pairs with similarity scores above a threshold $\tau_1 = 0.6$ are retained for further validation. As a result of this filtering process, 225,987 pairs are retained. The distribution of similarity scores (see Appendix~\ref{fig:pair_scoring}) shows that most pairs cluster in the higher score range, indicating that a large portion of translations preserve core functionality. Based on this distribution, we set $\tau_1 = 0.6$ to retain pairs with relatively high similarity while filtering out low-quality translations.

\noindent\textbf{AST-based Similarity Scoring:} Next, we perform a back-translation procedure to further verify the consistency. The validated COBOL program is translated back into Java using an LLM (Translator), producing a back-translated Java version. We then compare the original Java program with the back-translated Java program using an AST-based similarity scoring, which consists of two steps: AST normalization and similarity measurement.
First, to reduce noise introduced by superficial naming differences, we normalize both Java programs by abstracting identifier names. We use the Spoon framework\footnote{https://github.com/INRIA/spoon/} \citep{pawlak:hal-01169705} to parse each program into its abstract syntax tree (AST) representation and systematically rename variables, method parameters, and local identifiers into a canonical form. This process ensures that the comparison focuses on structural properties rather than changes in identifier names.
After normalization, we compute the similarity between the original and back-translated Java programs using CodeBERTScore \citep{zhou2023codebertscore}, which leverages contextual embeddings from a pre-trained model (CodeBERT) and computes token-level cosine similarity between the two programs. Pairs are retained only if their similarity scores exceed the threshold $\tau_2 = 0.7$. 
 We further analyze the effect of the normalization step in Appendix~\ref{fig:codebert}. Based on this distribution, where most pairs cluster above 0.7, we set $\tau_2 = 0.7$ to retain pairs with high similarity while filtering out inconsistent translations. 



After semantic validation, the filtered COBOL programs are used as inputs to the instruction generation stage (Stage 3), where we construct problem descriptions and standardized instruction–solution pairs following the same procedure described in Section~\ref{sec:source_code}.
Through this multi-stage pipeline, we obtain a large synthetic dataset consisting of Java–COBOL translation pairs, COBOL–Java translation pairs, and description–code instruction pairs, as summarized in Table~\ref{tab:training_data}.

\subsubsection{Source 3: COBOL and Mainframe Knowledge Sources}
\label{sec:document}

In addition to source code, we curate a corpus of textual resources covering COBOL and mainframe knowledge. These sources include licensed textbooks, tutorials, and technical websites that describe COBOL syntax, data structures, file systems, and execution environments. To ensure high content quality, we focus on explanatory and instruction-oriented materials, while excluding community-driven conversational sources (e.g., Stack Overflow, discussion forums, and mailing lists), which often emphasize problem-specific fixes rather than systematic domain knowledge.

For textbook sources, we obtain licensed PDF copies and extract text using the \texttt{pdftotext} \footnote{https://github.com/jalan/pdftotext} library. The extracted content is subsequently normalized and filtered to remove corrupted pages and non-informative segments.

For web-based sources, we implement a custom content extraction pipeline tailored to technical documents. We extract the main textual content by pruning the document structure and removing non-informative elements such as navigation menus, advertisements, sidebars, footers, and boilerplate text. This process combines tag-based filtering with keyword-based heuristics to preserve code blocks, tables, and structured examples while discarding irrelevant content. To further improve data quality, we remove duplicated documents and normalize formatting to maintain paragraph structure and code–text alignment.

After cleaning and filtering, the resulting documentation corpus consists of 18,498 high-quality text files, comprising approximately 37 million tokens.
Rather than directly using raw documentation as training data, we transform the corpus into instruction-style data for two main reasons. First, generating derived question–answer pairs helps mitigate potential copyright concerns by avoiding direct reproduction of proprietary or licensed content. Second, prior work shows that instruction-style data is more effective for training LLMs, whereas raw documentation often contains redundant or non-instructional information \citep{abdin2024phi,rowberry2025value}.

Following the pipeline shown in Figure~\ref{fig:code_pipeline}, we segment the extracted corpus into paragraphs and use an LLM (Synthesizer) to generate question–answer pairs for each segment. This process enables the model to learn underlying concepts and usage patterns rather than memorizing raw text. In total, we obtain 153,415 question–answer pairs. Combined with the curated source code datasets, this data provides contextual information, enabling \modelname to better capture COBOL-specific concepts and practices beyond the source code information.

We summarize the statistics of the constructed fine-tuning dataset across all sources and instruction formats in Table~\ref{tab:training_data}. 

\begin{table*}[]
\renewcommand{\arraystretch}{1.2}
\caption{The COBOL fine-tuning dataset across data sources and instruction formats. }
\centering
\begin{tabular}{l|l|c|c}
\hline
Data Source                           & Instruction Format     & Token Count & \# Instances \\ \hline
GitHub Repositories                   & Description—Code & 38.4M        & 31,492       \\ \hline
\multirow{3}{*}{Synthetic COBOL} & COBOL-Java             & 206M         & 173,042      \\ \cline{2-4} 
                                        & Java-COBOL             & 170M         & 173,042      \\ \cline{2-4} 
                                      & Description—Code & 230M         & 172,759      \\ \hline
COBOL Knowledge Sources                       & Question-Answer        & 241M        & 153,415      \\ \hline
\end{tabular}
\label{tab:training_data}
\end{table*}

\subsection{Instruction Tuning Details}
\label{sec:instruction}
\noindent\textbf{Model Architecture and Base Model Selection.} \modelname is fine-tuned on top of the Qwen2.5-Coder model family \citep{hui2024qwen2}, the leading code LLM \footnote{https://huggingface.co/spaces/bigcode/bigcode-models-leaderboard}. Qwen2.5-Coder is pretrained on a mixture of programming languages and natural language, with architectural design choices optimized for code understanding and generation, including long-context support and strong instruction-following capabilities. These properties make it a suitable foundation for the adaptation to legacy programming languages such as COBOL. 
Rather than introducing task-specific architectural changes, our approach focuses on language specialization through fine-tuning, similar to the previous work \citep{wei2023magicoder}. By retaining the base architecture, we ensure stable training, reproducibility, and fair comparison with existing code LLMs. This also allows us to isolate the effects of COBOL-adaptive data curation on downstream performance.
Unless otherwise specified, all experiments are conducted using the 7B and 14B parameter versions of Qwen2.5-Coder. We refer to the resulting variants of the COBOL-adapted model as \modelnamenospace, which inherit the general reasoning capabilities of the base model while acquiring specialized knowledge of COBOL syntax and mainframe conventions. 

\noindent\textbf{Training Details.}
We fine-tune all LLMs  with a maximum sequence length of 4,096 tokens. Training is performed with the AdamW optimizer \cite{kinga2015method}, using $\beta_{1}=0.9, \beta_{1}=0.95, \epsilon=10^{-8}$
, and a weight decay of 0.1. We adopt a cosine learning rate schedule with a linear warm-up of 1,000 steps, followed by a smooth decay to one-thirtieth of the peak learning rate.
For \modelnamenospace-7B, the learning rate is set to $2\times 10^{-5} $, while for \modelnamenospace-14B, we use a lower learning rate $1\times 10^{-5}$
 to ensure training stability at a larger scale. All versions are trained with a global batch size of 2,048 samples, with inputs packed into sequences of fixed length to maximize throughput. No dropout is applied during fine-tuning. LLMs are trained until convergence or early stopping based on validation performance.

\section{Evaluation}
Our evaluation aims to answer the research questions below:

\begin{description}
    \item[RQ1:] How well do \modelname and existing LLMs generate compilable and correct COBOL code?
    \item [RQ2:] How effective is \modelname at translating between COBOL and Java in both directions on \benchmarknamenospace?
    \item[RQ3:] How do experienced COBOL developers evaluate the quality and practical usefulness of LLM-generated code in realistic development scenarios?
\end{description}

\subsection{Experiment Setup}

\subsubsection{Baseline Models}
We consider a wide range of baseline models, including recent high-performing open-source code LLMs such as DeepSeek-Coder \citep{guo2024deepseek}, CodeGemma \citep{team2024codegemma}, CodeLlama \citep{roziere2023code}, StarCoder2 \citep{lozhkov2024starcoder}, Qwen2.5-Coder \citep{hui2024qwen2}, and DeepSeek-R1-Distill-Qwen  \citep{guo2025deepseek}. We additionally evaluate Mainframer \citep{noauthor_bloop_nodate}, the state-of-the-art LLM specifically designed for COBOL code generation. To investigate the effectiveness of general-purpose LLMs, we also evaluate several variants of GPT models: GPT-oss \citep{gptoss}, GPT-4 \citep{gpt4}, and GPT-4o \citep{gpt4o}.


\subsubsection{Metrics}
We assess the correctness of both generated and translated code using two metrics, including the Compilation Success Rate (CSR) and Pass@1.

\noindent\textbf{CSR:} It measures the proportion of generated solutions that compile successfully. 
We use the corresponding compiler for different languages  (e.g., \textit{GnuCOBOL}—version 2.0.0 for COBOL and \textit{javac}—version 17.0.18 for Java).

\noindent\textbf{Pass@1:} Pass@1 evaluates functional correctness by measuring the percentage of tasks for which the model’s first generated solution passes all test cases.


\subsubsection{Implementation Details}
We evaluate 11 LLMs for our experiments. All experiments on the open-source models, \modelnamenospace, and GPT-oss-120B are conducted on our local machine under the same computational settings. 
We use the official API interface \citep{gpt4,gpt4o} to run GPT-4 and GPT-4o. 
The hyperparameters for the evaluation process are set as follows: $temperature=0.0$ and $n=1$, which means that we only consider LLMs’ first candidates
for evaluation. 
All other hyperparameters are kept by default.
To mitigate the impact of LLM randomness on result reliability, following prior works \citep{liu2023your,yan2023codetransocean}, we repeat the
experiments three times under the same settings and report the averaged results. All evaluations are performed in a zero-shot setting.

\subsubsection{Benchmarks} Next, we describe the benchmarks used in this study to evaluate COBOL code generation and code translation.
 
\noindent\textbf{Benchmarks for Code Generation. }To evaluate COBOL-related capabilities of LLMs, we employ two established COBOL code generation benchmarks—COBOLEval \citep{noauthor_bloop_nodate} and COBOLCodeBench \citep{cobolcodebench}. COBOLEval is the first benchmark specifically designed for COBOL code generation. It consists of 146 programming tasks manually translated from the widely used HumanEval benchmark \citep{chen2021codex}, where each task provides a COBOL function signature and a natural language specification. 
COBOLCodeBench further increases evaluation difficulty by adapting 46 tasks from BigCodeBench-Hard \citep{zhuo2024bigcodebench} to COBOL, which emphasizes realistic legacy programming scenarios commonly found in enterprise systems, such as financial computations, structured data processing, and report generation. 


\label{sec:code_translation}
\renewcommand{\arraystretch}{1.2}
\begin{table*}[]
\centering
\footnotesize
\caption{Comparison of COBOL-related benchmarks used in our work. Gen denotes COBOL code generation,
C2J stands for COBOL-to-Java translation, and J2C represents Java-to-COBOL translation.}
\begin{tabular}{c|c|c|cc}
\hline
\multirow{2}{*}{Dataset name} & \multirow{2}{*}{Source Language} & \multirow{2}{*}{Task Category} & \multicolumn{2}{c}{Dataset Size}          \\ \cline{4-5} 
                              &                                  &                                & \multicolumn{1}{l|}{Instances} & Unit     \\ \hline
COBOLEval                     & Python                           & Gen                            & \multicolumn{1}{c|}{146}       & problems \\ \hline
COBOLCodeBench                & Python                           & Gen                            & \multicolumn{1}{c|}{46}        & problems \\ \hline
\benchmarkname & COBOL, Java                      & C2J, J2C                       & \multicolumn{1}{c|}{143}       & pairs    \\ \hline
\end{tabular}
\label{tab:compare_bench}
\end{table*}

\noindent\textbf{Constructing the \benchmarkname Benchmark for Code Translation.} 
To the best of our knowledge, there is currently no benchmark curated specifically for COBOL code translation.
Existing benchmarks, such as COBOLEval \citep{noauthor_bloop_nodate}, provide only COBOL function signatures and docstrings as model input, along with unit tests for evaluation, but do not include reference COBOL implementations. As a result, they are not suitable for translation tasks that require COBOL source code as input (e.g., COBOL-to-Java translation).
To address this limitation, we construct COBOL programs for HumanEval's tasks using a vibe-coding–inspired workflow, in which LLMs generate candidate COBOL programs that are subsequently refined through repeated prompting and manual correction. As a previous study shows that AI-generated code can be low quality \citep{fan2023automated}, all produced programs are manually reviewed and validated for compilability and functional consistency. Due to fundamental mismatches between some HumanEval
 tasks and COBOL structure, not all problems can be faithfully implemented; in total, 143 of the 164 tasks  are deemed suitable for COBOL implementation, meaning they can be compiled and successfully pass all the test cases. 
 HumanEval is originally designed for Python, while HumanEval-X \citep{zheng2023codegeex} extends it to multiple programming languages by providing implementations, including Java. Building on this extension, we construct our benchmark using Java solutions from HumanEval-X while adapting task specifications to ensure compatibility with COBOL.
Motivated by the practical demand for migrating legacy systems to modern platforms \citep{hans2025automated,sneed2010migrating}, particularly from COBOL to Java, we introduce \benchmarknamenospace, a COBOL–Java code translation benchmark. 
This benchmark enables systematic evaluation of translation performance in both directions. In our experiments, we report results for COBOL-to-Java translation, reflecting real-world modernization scenarios, as well as Java-to-COBOL translation, which tests models’ ability to synthesize idiomatic legacy code from modern language inputs. Table~\ref{tab:compare_bench} summarizes the key differences between \benchmarkname and the two other benchmarks used in this work in terms of \emph{source languages} (i.e., the language of reference solutions or the source input in translation tasks), supported task types, and dataset size. Unlike prior datasets, which focus primarily on code generation, \benchmarkname additionally supports bidirectional COBOL–Java translation, enabling more comprehensive evaluation of modernization scenarios.


\section{Evaluation Results}
\subsection*{RQ1: Comparison of \modelname with Existing LLMs for COBOL code generation}

\begin{table}[]
\renewcommand{\arraystretch}{1.2}
\caption{Performance of various LLMs on COBOLEval and COBOLCodeBench. \textbf{Bold} stands for the best in the block; \textbf{\underline{Bold and Underlined}} denotes the overall best.}
\centering
\begin{tabular}{l|rr|rr}
\hline
\multirow{2}{*}{Model}                               & \multicolumn{2}{c|}{COBOLEval}                                                    & \multicolumn{2}{c}{COBOLCodeBench}                                               \\ \cline{2-5} 
                                                     & CSR                                     & Pass@1                                  & CSR                                     & Pass@1                                 \\ \hline
DeepSeek-Coder 6.7B                                  & 14.98                                   & 1.37                                    & 0                                       & 0                                      \\
CodeGemma 7B                                         & 0                                       & 0                                       & 0                                       & 0                                      \\
CodeLlama 7B                                         & 0                                       & 0                                       & 0                                       & 0                                      \\
Mainframer 7B                                        & 69.17                         & 6.16                                    & 0                                       & 0                                      \\
Qwen2.5-Coder 7B                                     & 10.27                                   & 0.68                                    & 0                                       & 0                                      \\
DeepSeek-R1-Distill-Qwen-7B                          & 0                                       & 0                                       & 0                                       & 0                                      \\
StarCoder2 7B                                        & 0                                       & 0                                       & 0                                       & 0                                      \\
\textbf{\modelnamenospace-7B (Ours)}  & \textbf{73.80}                                   & \textbf{44.70}                          & \textbf{13.04}                          & 0                                      \\ \hline
CodeLlama 13B                                        & 3.40                                     & 0.68                                    & 0                                       & 0                                      \\
Mainframer 13B                                       & 62.24                                   & 11.64                                   & 0                                       & 0                                      \\
Qwen2.5-Coder 14B                                    & 12.32                                   & 2.74                                    & 0                                       & 0                                      \\
DeepSeek-R1-Distill-Qwen-14B                         & 0                                       & 0                                       & 0                                       & 0                                      \\
StarCoder2 15B                                       & 0                                       & 0                                       & 0                                       & 0                                      \\
DeepSeekCoder-V2 16B                                 & 10.27                                   & 1.37                                    & 0                                       & 0                                      \\
\textbf{\modelnamenospace-14B (Ours)} & \textbf{\underline{73.95}} & \textbf{\underline{49.33}} & \textbf{\underline{26.09}} & \textbf{\underline{4.35}} \\ \hline
GPT-oss-120B                                         & 19.17                                   & 4.11                                    & \textbf{17.39}                          & \textbf{2.17}                          \\
GPT-4                                                & 24.12                          & 15.75                                   & 13.04                                   & 0                                      \\
\textbf{GPT-4o}                                               & 41.80                                    & \textbf{16.40}                           & 13.04                                   & 0                                      \\ \hline
\end{tabular}

\label{tab:code_generation}
\end{table}
Table \ref{tab:code_generation} shows the evaluation results of various LLMs on two COBOL code generation benchmarks, COBOLEval and COBOLCodeBench, where the ``CSR'' columns represent compilation success rate (CSR), and the ``Pass@1'' columns represent the values for the Pass@1 metric. 
As shown in the results, GPT variants and code-oriented LLMs, despite strong performance on modern language benchmarks, consistently struggle with COBOL.
Most open-source baselines, including CodeGemma, CodeLlama, StarCoder2, and DeepSeek-R1-Distill-Qwen, achieve 0\% CSR and 0 Pass@1 on both COBOLEval and COBOLCodeBench. Meanwhile, stronger Code LLMs such as DeepSeek-Coder 6.7B and Qwen2.5-Coder 7B achieve only 14.98\% and 10.27\% CSR, respectively, with Pass@1 below 1.5 on COBOLEval, and fail to generate any valid programs on COBOLCodeBench (i.e., the more challenging benchmark). 
The GPT variants show better results: GPT-4 achieves 24.12\% CSR and 15.75 Pass@1 on COBOLEval, while GPT-4o improves CSR to 41.8\% but reaches only 16.4 Pass@1, and neither model exceeds 2.17 Pass@1 on COBOLCodeBench. 
In contrast, \modelname significantly outperforming all baselines
in both syntactic correctness (CSR) and functional correctness (Pass@1). Moving from 7B to 14B parameters consistently improves performance of \modelnamenospace, with the 14B variant achieving the highest performance on both benchmarks. \modelnamenospace-7B reaches 73.80\% CSR and 44.70 Pass@1 on COBOLEval, outperforming all 7B baselines and exceeding GPT-4o’s Pass@1 by nearly 30. \modelnamenospace-14B achieves 72.95\% CSR and 49.33 Pass@1 on COBOLEval, and is the only model to obtain non-trivial functional correctness on COBOLCodeBench with 26.09\% CSR and 4.35 Pass@1. These results indicate that both COBOL-specific training and model scaling are critical for COBOL code generation. 

\subsection*{RQ2: Effective of \modelname in translating between COBOL and Java}

\begin{table}[]
\caption{Performance of various LLMs on the \benchmarkname benchmark. \textit{C2J} denotes COBOL-to-Java translation, while \textit{J2C} indicates Java-to-COBOL translation. \textbf{Bold} stands for the best in the block; \textbf{\underline{Bold and Underlined}} denotes the overall best.}
\centering
\begin{tabular}{l|rr|rr}
\hline
\multirow{2}{*}{Model}                               & \multicolumn{2}{c|}{C2J}                                    & \multicolumn{2}{c}{J2C}                                     \\ \cline{2-5} 
                                                     & CSR                                     & Pass@1                                  & CSR                                     & Pass@1                                  \\ \hline
DeepSeek-Coder 6.7B                                  & 88.11                                   & 63.64                                   & 0                                       & 0                                       \\
CodeGemma 7B                                         & 76.22                                   & 48.25                                   & 0                                       & 0                                       \\
CodeLlama 7B                                         & 76.92                                   & 29.37                                   & 0                                       & 0                                       \\
Mainframer 7B                                        & 5.59                                    & 1.39                                    & 0                                       & 0                                       \\
Qwen2.5-Coder 7B                                     & 14.68                                   & 10.47                                   & 0                                       & 0                                       \\
DeepSeek-R1-Distill-Qwen-7B                          & 83.21                                   & 55.94                                   & 0                                       & 0                                       \\
StarCoder2 7B                                        & 0                                       & 0                                       & 0                                       & 0                                       \\
\textbf{\modelnamenospace-7B (Ours)}  & \textbf{97.90}                          & \textbf{81.81}                          & \textbf{63.64}                          & \textbf{27.27}                          \\ \hline
CodeLlama 13B                                        & 83.21                                   & 48.95                                   & 0                                       & 0                                       \\
Mainframer 13B                                       & 62.23                                   & 37.06                                   & 0                                       & 0                                       \\
Qwen2.5-Coder 14B                                    & 8.39                                    & 3.50                                    & 0                                       & 0                                       \\
DeepSeek-R1-Distill-Qwen-14B                         & 70.63                                   & 60.13                                   & 0                                       & 0                                       \\
StarCoder2 15B                                       & 39.36                                   & 18.88                                   & 0                                       & 0                                       \\
DeepSeekCoder-V2 16B                                 & 95.10                                   & 75.52                                   & 0                                       & 0                                       \\
\textbf{\modelnamenospace-14B (Ours)} & \textbf{97.90}                          & \textbf{83.91}                          & \textbf{\underline{72.03}} & \textbf{\underline{34.93}} \\ \hline
GPT-oss-120B                                         & \textbf{\underline{98.60}} & \textbf{\underline{89.51}} & 5.38                                    & \textbf{3.93}                                    \\
GPT-4                                                & 94.40                                    & 72.73                                   & \textbf{5.45}                                    & 1.73                                    \\
GPT-4o                                               & 97.20                                   & 85.31                                   & 4.36                                    & 2.18                                    \\ \hline
\end{tabular}
\label{tab:code_translation}
\end{table}

Table \ref{tab:code_translation} presents the performance of various LLMs on the \benchmarkname benchmark for COBOL-to-Java and Java-to-COBOL translation. For the COBOL-to-Java translation task, our specialized LLMs, \modelnamenospace, consistently outperform all other open-source LLMs, achieving CSR scores of 97.90\% for both sizes and Pass@1 of 81.11 and 83.91, respectively. This represents a substantial improvement in terms of performance over code LLMs such as DeepSeek-Coder, CodeLlama, and StarCoder2, highlighting the significant benefit of model specialization for COBOL code translation. Notably, even larger general-purpose LLMs like GPT-oss-120B achieve slightly higher CSR and Pass@1 (98.6\% and 89.51), but at a much higher model scale, confirming that smaller, language-specialized LLMs can reach competitive performance. For the Java-to-COBOL direction, which is considerably more challenging, most LLMs fail to produce any successful compilations, with CSR and Pass@1 scores at 0. In contrast, \modelname achieves 63.64\% CSR and 27.27 Pass@1 for the 7B version and 72.03\% CSR and 34.93 Pass@1 for the 14B version. This demonstrates that our LLMs are effective in handling the reverse translation task, where general-purpose and even larger LLMs perform poorly. GPT-based LLMs show limited success (4–5\% CSR, Pass@1 below 4), further demonstrating the need for COBOL-specific training to tackle complex bidirectional translations.



\subsection*{RQ3: Developers' Feedback on the Quality and Usability of LLM-generated Code}
To complement our quantitative evaluation of LLMs for COBOL code generation and translation, we conduct a survey by inviting experienced COBOL developers. 
 The survey is designed to assess the practical utility of LLM-generated code from the perspective of practitioners actively working in this domain. 
We first describe the survey design and participant recruitment in Section \ref{sec:survey_design}. We then report observations derived from task-level rankings in Section \ref{sub:task} and gain insight from participants' qualitative feedback in Section \ref{sub:insight}.

\subsubsection{Survey Design and Participants.}
\label{sec:survey_design}

\begin{table*}[t]
\renewcommand{\arraystretch}{1.2}
\centering
\caption{Overview of tasks used in the practitioner survey.
\textit{Gen} denotes COBOL code generation, \textit{C2J} denotes COBOL-to-Java translation, and \textit{J2C} denotes Java-to-COBOL translation.}
\begin{tabular}{c|c|l|l}
\hline
Task ID & Category & Task Title & Difficulty \\ \hline
1  & Gen & COPYBOOK-based transaction aggregation & Simple \\
2  & Gen & Line-sequential file processing loop & Simple \\
3  & Gen & Record validation and error routing & Moderate \\
4  & Gen & Table search with OCCURS and SEARCH & Moderate \\
5  & Gen & DB2 cursor fetch and interest computation & Complex \\ \hline
6  & C2J & Salary bonus calculation & Simple \\
7  & J2C & Batch file line counting & Simple \\
8  & C2J & Customer record validation logic & Moderate \\
9  & J2C & Transaction array aggregation & Moderate \\
10 & J2C & Stateful processing with business-rule termination & Complex \\ \hline
\end{tabular}
\label{tab:survey_tasks}
\end{table*}
Recruiting participants is inherently challenging given COBOL's status as a legacy language with a shrinking developer community; consequently, we invited three professionals from our network with industrial COBOL experience: one with 1–3 years, one with 5–10 years, and one with more than 10 years of experience maintaining and developing production COBOL applications.

\begin{figure*}[ht]
\centering
\includegraphics[width=\textwidth]{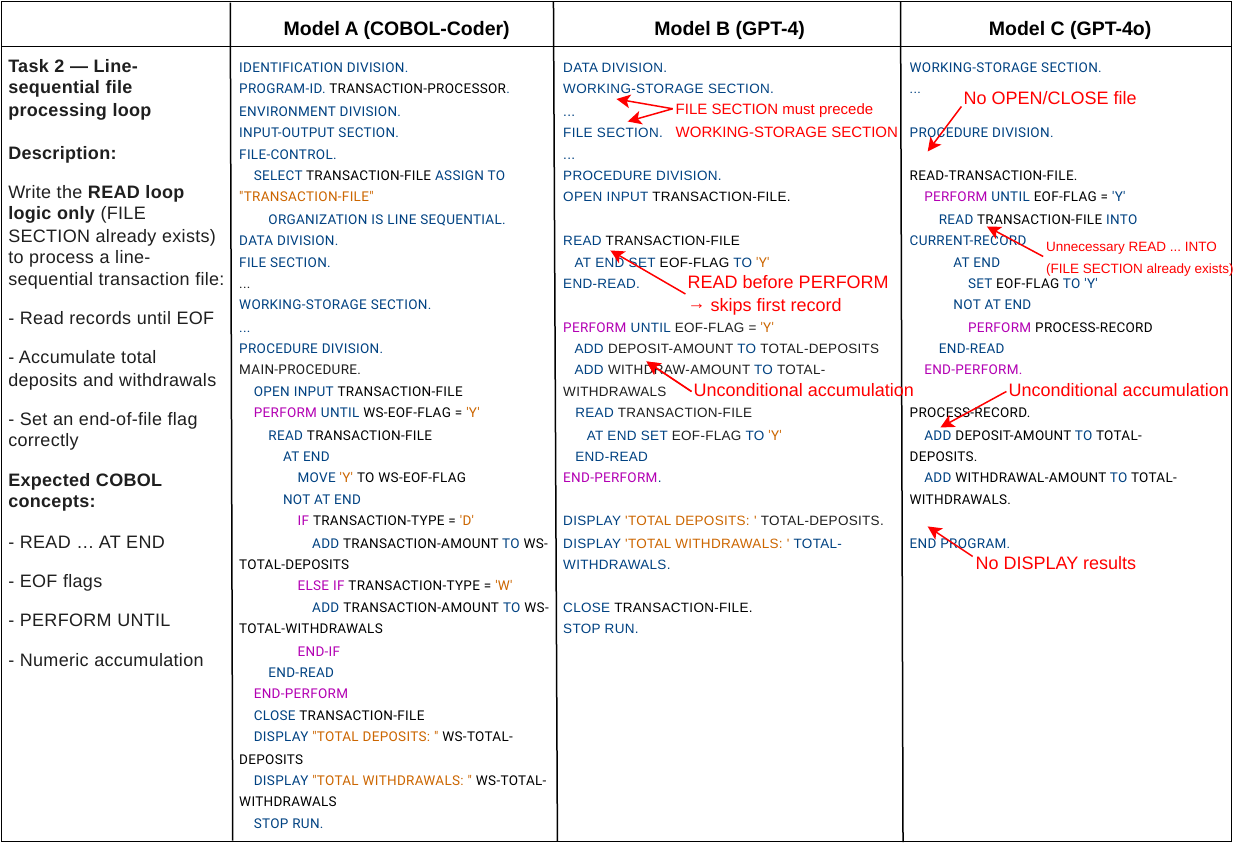}
\caption{Example of a survey task and corresponding COBOL code generated by three models. \textcolor{red}{Red arrows} highlight errors in code snippets.}
\label{fig:example}
\end{figure*}

The survey consisted of 10 tasks covering COBOL code generation and bidirectional translation between COBOL and Java, designed to reflect common and practical programming scenarios such as file processing, table operations, and business rule validation, ranging from relatively simple to more complex cases, as shown in Table \ref{tab:survey_tasks}. For each task, participants were presented with three LLM-generated solutions, labeled as Model A, B, and C, and asked to rank them according to three criteria: (1) functional correctness, (2) code readability, and (3) adherence to conventional COBOL coding practices. Figure \ref{fig:example} illustrates an example survey task used in our study together with COBOL code generated by three models for comparison. To reduce potential bias, the identities of the underlying models were anonymized during evaluation. The solutions were generated by \modelnamenospace-14B, and widely used general-purpose LLMs, GPT-4 and GPT-4o, respectively.
Participants completed the survey individually by ranking the LLM-generated solutions for each task. At the end of the survey, they were asked to provide feedback on their experience, including the perceived usefulness, clarity, and potential limitations of the generated code. 

\subsubsection{Insights from Practitioner Rankings Across Tasks}
\label{sub:task}

\begin{table}[]
\renewcommand{\arraystretch}{1.2}
\caption{Rankings by three experienced COBOL developers (P1–P3) across ten tasks. \textit{Gen} denotes COBOL code generation, \textit{C2J} denotes COBOL-to-Java translation, and \textit{J2C} denotes Java-to-COBOL translation.
\textit{Model A:} \modelnamenospace-14B, \textit{Model B:} GPT-4, and \textit{Model C:} GPT-4o.
Model identities were anonymized during evaluation.
For each task, the ranking order reflects the participant’s preference from best to worst.
“No preference” means that a participant assigned equal rank to all three LLMs, indicating no observable quality difference.}
\centering
\begin{tabular}{c|c|c|c|c}
\hline

   Task ID     & Category                      & P1                          & P2                          & P3                          \\ \hline
1  & Gen     & B - A - C & C - B - A & A - B - C \\
2  & Gen     & A - C - B & A - C - B & A - C - B \\
3  & Gen     & B - C - A & A - B - C & No preference         \\
4  & Gen    & A - B - C & A - C - B & No preference         \\
5  & Gen    & A - B - C & No preference         & C - A - B \\
6  & C2J & B - C - A & B - C - A & A - B - C \\
7  & J2C & A - B - C & A - C - B & A - B - C \\
8  & C2J & B - C - A & C - B - A & C - B - A \\
9  & J2C  & A - B - C & C - A - B & A - B - C \\
10 & J2C & A - C - B & A - C - B & A - C - B \\ \hline
\end{tabular}
\label{tab:survey_response}
\end{table}

\noindent\textbf{Results Discussions.} We analyze participant rankings across the ten evaluation tasks, grouped by task category, as shown in Table \ref{tab:survey_response}. 

\noindent\textbf{COBOL Code Generation:} For COBOL code generation tasks (Tasks 1–5), \modelname (Model A) is frequently ranked first or tied for first across participants.
In simpler tasks, participants often reported that all LLMs produced functionally similar solutions. In these cases, general-purpose LLMs (Models B and C) were occasionally ranked higher due to their use of explanatory comments and more verbose descriptions. However, participants noted that the underlying code quality across models was largely comparable for these tasks, indicating that \modelname remains competitive even when not ranked first. 
For tasks of moderate or higher complexity, \modelname was preferred due to its program structure, use of batch-oriented patterns, and adherence to conventional COBOL coding practices.

\noindent\textbf{COBOL-to-Java Translation:}
In COBOL-to-Java translation tasks, \modelname remained competitive but did not consistently dominate. Participants often preferred GPT-4 or GPT-4o when evaluating Java, reflecting their stronger performance for high-resource, modern programming languages. Nevertheless, \modelnamenospace’s translations were generally assessed as functionally correct and structurally sound.

\noindent\textbf{Java-to-COBOL Translation:}
In contrast, \modelname clearly outperforms GPT-4 and GPT-4o across all Java-to-COBOL translation tasks (Tasks 7, 9, and 10). Participants consistently ranked Model A (\modelnamenospace) first. These results align with our experimental findings in Section \ref{sec:code_translation}, indicating that \modelname is particularly effective when translating a modern language into legacy COBOL code.

\subsubsection{Insights from Practitioner's Qualitative Feedback}
\label{sub:insight}

\noindent\textbf{Understanding Program Intent and COBOL Semantics: }
Participants consistently reported differences among the three LLMs. 
Rather than judging outputs purely by correctness, developers evaluated whether the generated code reflected an understanding of the program logic, how the code would be executed, and how it would evolve in a real system.
\modelname (Model A) was described as more pragmatic and production-oriented, whereas GPT-4 (Model B) and GPT-4o (Model C) exhibited weaknesses either in correctness or structural appropriateness.
Participants summarized this contrast as:

\begin{quote}
\textit{
“Model A felt pragmatically strong and production-oriented. Model B was inconsistent and often incorrect, while Model C understood the concepts but tended to overengineer the solution.”}  (\textit{P1})

\textit{“Sometimes the differences between models were very clear, and sometimes two models were very similar—one might be right in structure but wrong in details, while the other was the opposite.”}
(\textit{P2})
\end{quote}

\noindent\textbf{COBOL Awareness and Program Structure:}
Across responses, \modelname (Model A) was most frequently identified as the model that best understood COBOL-specific conventions, enterprise programming style, and program organization. All three participants selected Model A as the most COBOL-aware model.

\begin{quote}

\textit{“Model A felt the most COBOL-aware overall.”} (\textit{P1})

\textit{“Model A was closer to a typical batch-style program, especially for file processing. The other models felt more general.”} (\textit{P3})

\end{quote}

While some tasks resulted in similar code quality across LLMs, \modelname was described as more consistent in preserving program structure, whereas other LLMs \textit{“sometimes lost the structure of the program”} \textit{(P2)}.

\noindent\textbf{Productivity and Developer Workflow:}
From a workflow perspective, \modelname was perceived as offering productivity benefits, particularly by reducing cognitive load during early development stages. Participant 2 described it as shortening the path from an initial idea to a reasonably correct draft, which is especially valuable in COBOL development and migration scenarios where boilerplate logic and legacy constraints are common.

\begin{quote}
    \textit{“Model A reduced cognitive load. It shortened the path from idea $\to$ correct implementation, which is exactly what improves efficiency when working with COBOL, Java migrations, and legacy logic.”} \textit{(P2)}

\end{quote}

However, participants framed this benefit within a human-in-the-loop workflow. LLMs were viewed as accelerators rather than autonomous agents, comparable to junior developers producing first drafts. 
In contrast, they found it difficult to identify clear productivity advantages for GPT-4 and GPT-4o in practical settings, particularly when their outputs required substantial restructuring and correction.

\noindent\textbf{Requirements for Real-World Adoption:}
When discussing requirements for real-world usage, participants converged on several concrete expectations. First, the elimination of non-compilable constructs was considered a prerequisite. Second, predictable handling of operational edge cases—such as file status checks, handling of SQLCODE (i.e., a variable that stores the most recently executed SQL statement), and restartability (i.e., the ability to resume a failed batch job from an intermediate ``checkpoint'' instead of the start)—was viewed as essential for enterprise deployment. Third, consistent adherence to organizational copybooks, data conventions, and coding standards was identified as a major barrier to adoption. Fourth,  understanding  copybooks, shared data layouts, job control language (JCL), which is used in mainframe batch environment, and cross-program dependencies.

\modelname was widely regarded as the closest to meeting these requirements, particularly as a first-draft generator. However, participants emphasized that deeper integration with existing mainframe ecosystems and stronger guarantees around safety and compatibility would be necessary before LLMs could be used routinely in production environments.

Overall, the qualitative feedback reinforces the quantitative findings: \modelname is perceived as substantially more aligned with COBOL development practices than the baseline LLMs. At the same time, the feedback highlights that practical adoption depends not only on model accuracy but also on trust, predictability, and integration into established mainframe workflows.

\section{Related Work}
\noindent\textbf{Large Language Models for Code Generation.}
LLMs have achieved strong performance across a wide range of code-related tasks. Early work such as Codex \citep{chen2021codex} demonstrated the effectiveness of large-scale pretraining on mixed natural language and source code corpora. Subsequent LLMs, including CodeLlama \citep{roziere2023code}, StarCoder \citep{li2023starcoder,lozhkov2024starcoder}, DeepSeek-Coder \citep{guo2024deepseek}, and commercial systems such as GPT-4 \citep{gpt4} and Claude Sonnet \citep{claude}, further improved reasoning and code synthesis capabilities through instruction tuning and reinforcement learning. Despite these advances, most existing code LLMs are optimized for high-resource, modern programming languages such as Python and Java. Performance on legacy languages remains limited, particularly for languages with rigid syntactic constraints and domain-specific conventions \citep{dau2024xmainframe}. Recent studies have shown that general-purpose code LLMs struggle with long-range dependencies, strict formatting rules, and low-frequency language constructs, all of which are unique characteristics of COBOL programs \citep{noauthor_bloop_nodate,lei2025enhancing}.
Unlike prior approaches that treat code LLMs as general-purpose tools, our work demonstrates that targeted fine-tuning on curated domain-specific data consistently outperforms much larger general-purpose LLMs on legacy language tasks, highlighting the primacy of data quality over model scale in low-resource settings.

\noindent\textbf{Low Resource Languages and Domain-Specific Adaptation.}
Low Resource Programming Languages (LRPLs) are characterized by the limited availability of publicly accessible training data, which leads to systematic performance degradation for large language models compared to high-resource programming languages (HRPLs) such as Python and Java \citep{joel2024survey}. This disparity is consistently observed across modern code generation benchmarks. For example, results on the MultiPL-E \citep{cassano2023multipl} benchmark show that state-of-the-art LLMs achieve substantially higher Pass@1 scores on HRPLs than on LRPLs, with performance gaps persisting across model families and scales \citep{joel2024survey}. These findings suggest that improvements in model architecture or parameter count alone are insufficient to close the gap for low-resource languages.

Domain-Specific Languages (DSLs) exhibit similar limitations. Although DSLs offer concise abstractions and improved expressiveness within specialized domains, their niche usage results in a lack of training data and limited benchmark \citep{joel2024survey,luo2025unlocking}. Prior studies report notable performance drops when LLMs are applied to DSLs with unique syntax, semantics that deviate from large-corpus programming languages \citep{joel2024survey}. 

COBOL shares core characteristics with both LRPLs and DSLs. While not niche in industrial impact, COBOL suffers from severe data scarcity in modern code corpora. As a result, general-purpose code LLMs frequently fail to generate compilable COBOL programs \citep{dau2024xmainframe}.
Insights from low-resource natural language processing provide a useful lens to address these challenges. Prior work shows that domain-adaptive pretraining and targeted fine-tuning can significantly improve performance in low-resource settings, even when monolingual data is limited \citep{yan2025adaft,alyami2025domain}. Techniques such as cross-lingual transfer, multilingual pretraining, and task-specific adaptation have been shown to yield more consistent gains than naïvely scaling model size \citep{lankford2023adaptmllm,song2025llm}. Recent studies extend these principles to code generation, showing that carefully curated domain-specific datasets and objectives are often more effective than generic pretraining for low-resource languages \citep{joel2024survey,luo2025unlocking}. Our work extends these insights by treating COBOL as a low-resource programming language and applying domain-adaptive fine-tuning strategies tailored to its syntactic and semantic properties.




\noindent\textbf{Benchmarks for Code and Legacy Languages.}
Code-related datasets have been widely developed to support empirical research across programming languages and software engineering tasks \citep{chen2021codex, austin2021program, zhuo2024bigcodebench}. However, legacy and low-resource programming languages, particularly COBOL, remain underrepresented, limiting reliable evaluation of LLMs for real-world legacy systems.

Several COBOL benchmarks have been proposed, including OpenCBS \citep{lee2022opencbs}, which uses public forums to construct a COBOL dataset for defect detection, and X-COBOL \citep{ali2023x}, which collects repositories from GitHub but relies on weak popularity-based quality filters \citep{borges2018s}. More recently, COBOLEval \citep{noauthor_bloop_nodate} and COBOLCodeBench \citep{cobolcodebench} adapt modern code benchmarks to COBOL, enabling execution-based evaluation for code generation, while MainframeBench \citep{dau2024xmainframe} evaluates broader mainframe knowledge but not program-level transformations. Recent work \citep{gandhi2024translation} leverages CodeNet \citep{puri2021codenet}, selecting problems that include both accepted COBOL and Java submissions to evaluate translation via execution-based testing. However, CodeNet originates from competitive programming platforms and is not designed around enterprise modernization scenarios

To address this gap, we introduce \benchmarknamenospace, a benchmark specifically designed for bidirectional COBOL–Java translation with testcases, enabling systematic evaluation of legacy–modern code translation tasks that remain largely absent from existing benchmarks.

\section{Threats to Validity}
\noindent\textbf{Construct Validity.}
In our data augmentation pipeline, we primarily rely on compiler feedback and LLMs to assess and filter generated COBOL programs. Compiler-based validation provides a strong and objective signal for syntactic correctness and structural well-formedness, which is particularly important for COBOL. However, this approach does not fully capture deeper semantic properties. Besides, relying on LLMs to select the most faithful instruction-response pairs and descriptions may also introduce bias due to their own training data and preferences. While we partially address these limitations through execution-based metrics (e.g., Pass@1) and bidirectional translation tasks that stress semantic preservation, our pipeline does not yet perform fine-grained semantic verification beyond these checks. Future work will explore richer semantic validation strategies, such as specification-based testing and expert-in-the-loop review, to further improve the quality and reliability of the curated training data.

\noindent\textbf{Internal Validity.}
A potential threat to internal validity arises from the scope of our dataset. In this work, we focus on file-level COBOL programs, treating each source file as an independent unit for generation, translation, and evaluation. This design choice simplifies benchmarking and enables controlled comparisons across LLMs, but it does not capture cross-file dependencies, shared copybooks, or inter-program control flow that are common in real-world mainframe systems. To mitigate this limitation, we construct our benchmarks and emphasize compilation- and execution-based validation, which already enforces many structural constraints. Nevertheless, repository-level reasoning remains an open challenge. In the future, we plan to extend our pipeline to repository-level settings, where LLMs must reason over multiple interconnected source files, shared data definitions, and system-wide conventions—an essential capability for large-scale legacy modernization.

\noindent\textbf{External Validity.}
Although our study includes multiple datasets covering code generation and bidirectional COBOL–Java translation, these benchmarks represent a subset of the diversity found in industrial mainframe environments. Real-world systems often involve proprietary libraries, organization-specific coding standards, and operational constraints that are not fully reflected in public datasets. To improve external validity, we complement evaluation with qualitative feedback from experienced COBOL developers, whose insights help contextualize model behavior in realistic development workflows. While the number of participants is necessarily limited due to the specialized expertise required, their feedback provides valuable signals about usability, structure, and production readiness. We will involve broader industrial collaboration and larger-scale evaluations to further validate the applicability of our approach across diverse enterprise settings.

\section{Reflection and Lessons Learned}
Based on our empirical study and survey, we discuss the main lessons learned from developing and evaluating LLMs for COBOL code generation and translation.

\noindent\textbf{Domain specialization is essential for legacy languages.} 
Our evaluation shows that Code LLMs struggle to generate high-quality COBOL code.
While modern code-oriented LLMs perform well on modern high-resource languages, they often fail to capture COBOL-specific syntax, structural conventions, and enterprise-oriented programming patterns. This suggests that COBOL is not only low-resource but also highly contextual, shaped by decades of operational practices. Domain adaptation—through targeted data curation, training, and evaluation—is therefore critical for reliable COBOL code generation and translation. 

\noindent\textbf{Legacy-to-modern and modern-to-legacy translations are fundamentally asymmetric.}
We observed a clear difference in difficulty between translating COBOL to Java and translating Java to COBOL. While general-purpose LLMs can often produce reasonable Java code from COBOL inputs, generating idiomatic and operationally correct COBOL remains substantially harder. This asymmetry reflects deeper differences in execution LLMs, data handling, and control-flow conventions between legacy and modern languages. As a result, translation tasks involving legacy targets require dedicated benchmarks and modeling strategies rather than symmetric treatment.

\noindent\textbf{LLMs are most effective as human-in-the-loop draft generators.} Experienced COBOL developers consistently viewed LLMs as productivity aids rather than autonomous developers. The most effective usage pattern resembled that of a junior developer producing an initial draft, which is then refined through compilation checks, testing, and code review. Productivity gains were most evident when the model reduced cognitive load and boilerplate effort, particularly in migration and maintenance scenarios. This highlights the importance of designing LLM-based tools to support assisted workflows instead of fully automated pipelines.

\noindent\textbf{Repository-level and system-level context are essential for real-world adoption.}
Our current evaluation focuses on single-file programs which aligns with existing benchmarks but does not fully reflect real mainframe systems. Practitioner feedback suggests that true production readiness depends on understanding COBOL intrinsics such as copybooks, shared data layouts, job control language (JCL), and cross-program dependencies. This insight points toward repository-level and system-level modeling as a critical next step for LLMs targeting enterprise COBOL environments.

\section{Conclusion}
This paper reports our experience in studying and improving the use of large language models for COBOL. We address a gap in existing research by combining domain-specific data curation, model adaptation, benchmark construction, and practitioner-centered evaluation for a legacy language that remains widely used in mission-critical systems.
In this paper, we present an automated data augmentation pipeline to construct the COBOL training data, introduce \benchmarknamenospace, a new benchmark for bidirectional COBOL–Java translation, and develop \modelnamenospace, a domain-adapted LLMs that substantially outperform general-purpose LLMs on both COBOL code generation and translation tasks. Our experiment results show that domain specialization and model scaling are both necessary to achieve meaningful gains on legacy languages.
We complemented these results with a qualitative study involving experienced COBOL developers working in realistic settings. Their feedback confirms that improvements measured by benchmarks translate into practical value: \modelname is perceived as more COBOL-aware and structurally reliable, especially for modernization and migration tasks. At the same time, the study highlights that human oversight and system-level context remain essential for real-world adoption.

\begin{acknowledgements}
This work is supported by the Natural Sciences and Engineering Research Council of Canada (NSERC) Discovery Grant
RGPIN-2024-04301.
The authors gratefully acknowledge the support and resources provided by Dr. Phong Xuan Nguyen of the FPT Software AI Center, Vietnam, which were essential to the completion of this work.
\end{acknowledgements}

\section*{Declarations}

\noindent\textbf{Funding:}  
Not applicable.

\noindent\textbf{Ethical Approval:}  
Not applicable.

\noindent\textbf{Informed Consent:}  
All participants received invitation from our network and have given their signed informed consent to participate.

\noindent\textbf{Author Contributions:}  

Conceptualization: Anh T. V. Dau, Shin Hwei Tan, Jinqiu Yang, Nghi D. Q. Bui, Anh Tuan Nguyen. 
Data Curation: Anh T. V. Dau. 
Methodology: Anh T. V. Dau, Shin Hwei Tan, Jinqiu Yang, Anh Tuan Nguyen. 
Experiment Analysis: Anh T. V. Dau, Shin Hwei Tan, Jinqiu Yang, Anh Tuan Nguyen. 
Supervision, supervising experiment design and execution: Shin Hwei Tan, Jinqiu Yang, Anh Tuan Nguyen. 
Writing – Original Draft: Anh T. V. Dau, Shin Hwei Tan, Jinqiu Yang. 
Writing – Review \& Editing: Anh T. V. Dau, Shin Hwei Tan, Jinqiu Yang

\noindent\textbf{Data and Code Availability Statement:}  
The benchmark and code used for our evaluation are publicly available: \url{https://github.com/COBOL-Coder/COBOL-Coder}. 

\noindent\textbf{Conflict of Interest:}  
The authors declare that they have no conflict of interest.

\noindent\textbf{Clinical Trial Number:}  
Not applicable.

\appendix
\section*{Appendix}
\bibliographystyle{spbasic}     
\bibliography{custom}

@String{Springer = "Springer-Verlag" }

@inproceedings{sneed2010migrating,
  title={Migrating from COBOL to Java},
  author={Sneed, Harry M},
  booktitle={2010 IEEE International Conference on Software Maintenance},
  pages={1--7},
  year={2010},
  organization={IEEE}
}

@inproceedings{hans2025automated,
  title={Automated Testing of COBOL to Java Transformation},
  author={Hans, Sandeep and Kumar, Atul and Yasue, Toshiaki and Ono, Kouichi and Krishnan, Saravanan and Sondhi, Devika and Satoh, Fumiko and Mitchell, Gerald and Kumar, Sachin and Saha, Diptikalyan},
  booktitle={Proceedings of the 33rd ACM International Conference on the Foundations of Software Engineering},
  pages={227--237},
  year={2025}
}

@book{litecky1974study,
  title={A study of errors, error-proneness, and error diagnosis of programming languages with special reference to cobol.},
  author={Litecky, Charles Roger},
  year={1974},
  publisher={University of Minnesota}
}

@inproceedings{fan2023automated,
  title={Automated repair of programs from large language models},
  author={Fan, Zhiyu and Gao, Xiang and Mirchev, Martin and Roychoudhury, Abhik and Tan, Shin Hwei},
  booktitle={2023 IEEE/ACM 45th International Conference on Software Engineering (ICSE)},
  pages={1469--1481},
  year={2023},
  organization={IEEE}
}

@article{zhou2023codebertscore,
  url = {https://arxiv.org/abs/2302.05527},
  author = {Zhou, Shuyan and Alon, Uri and Agarwal, Sumit and Neubig, Graham},
  title = {CodeBERTScore: Evaluating Code Generation with Pretrained Models of Code},  
  publisher = {arXiv},
  year = {2023},
  journal={   }
}

@article{pawlak:hal-01169705,
  TITLE = "{Spoon: A Library for Implementing Analyses and Transformations of Java Source Code}",
  AUTHOR = {Pawlak, Renaud and Monperrus, Martin and Petitprez, Nicolas and Noguera, Carlos and Seinturier, Lionel},
  JOURNAL = "{Software: Practice and Experience}",
  PUBLISHER = "{Wiley-Blackwell}",
  PAGES = {1155-1179},
  VOLUME = {46},
  URL = {https://hal.archives-ouvertes.fr/hal-01078532/document},
  YEAR = {2015},
  doi = {10.1002/spe.2346},
}

@misc{gpt4o,
        author={OpenAI},
	title = {Hello GPT 4o},
	url = {https://openai.com/index/hello-gpt-4o/},
	language = {en},
	urldate = {2025-02-16},
    year={2024}
}

@misc{gpt4,
        author={OpenAI},
	title = {GPT-4 is OpenAI’s most advanced system, producing safer and more useful responses},
	url = {https://openai.com/index/gpt-4/},
        year ={2023},
	language = {en},
	urldate = {2025-02-16},
}

@article{dau2024xmainframe,
  title={XMainframe: A Large Language Model for Mainframe Modernization},
  author={Dau, Anh TV and Dao, Hieu Trung and Nguyen, Anh Tuan and Tran, Hieu Trung and Nguyen, Phong X and Bui, Nghi DQ},
  journal={arXiv preprint arXiv:2408.04660},
  year={2024}
}

@article{roziere2023code,
  title={Code llama: Open foundation models for code},
  author={Roziere, Baptiste and Gehring, Jonas and Gloeckle, Fabian and Sootla, Sten and Gat, Itai and Tan, Xiaoqing Ellen and Adi, Yossi and Liu, Jingyu and Sauvestre, Romain and Remez, Tal and others},
  journal={arXiv preprint arXiv:2308.12950},
  year={2023}
}

@article{team2024codegemma,
  title={Codegemma: Open code models based on gemma},
  author={Team, CodeGemma and Zhao, Heri and Hui, Jeffrey and Howland, Joshua and Nguyen, Nam and Zuo, Siqi and Hu, Andrea and Choquette-Choo, Christopher A and Shen, Jingyue and Kelley, Joe and others},
  journal={arXiv preprint arXiv:2406.11409},
  year={2024}
}

@article{rowberry2025value,
  title={The value of books in the age of generative AI training data},
  author={Rowberry, Simon},
  journal={Convergence},
  pages={13548565251358020},
  year={2025},
  publisher={SAGE Publications Sage UK: London, England}
}

@article{abdin2024phi,
  title={Phi-4 technical report},
  author={Abdin, Marah and Aneja, Jyoti and Behl, Harkirat and Bubeck, S{\'e}bastien and Eldan, Ronen and Gunasekar, Suriya and Harrison, Michael and Hewett, Russell J and Javaheripi, Mojan and Kauffmann, Piero and others},
  journal={arXiv preprint arXiv:2412.08905},
  year={2024}
}

@article{guo2025deepseek,
  title={Deepseek-r1: Incentivizing reasoning capability in llms via reinforcement learning},
  author={Guo, Daya and Yang, Dejian and Zhang, Haowei and Song, Junxiao and Zhang, Ruoyu and Xu, Runxin and Zhu, Qihao and Ma, Shirong and Wang, Peiyi and Bi, Xiao and others},
  journal={arXiv preprint arXiv:2501.12948},
  year={2025}
}

@inproceedings{kinga2015method,
  title={A method for stochastic optimization},
  author={Kinga, Diederik and Adam, Jimmy Ba and others},
  booktitle={International conference on learning representations (ICLR)},
  volume={5},
  year={2015},
  organization={California;}
}

@article{li2023starcoder,
  title={Starcoder: may the source be with you!},
  author={Li, Raymond and Allal, Loubna Ben and Zi, Yangtian and Muennighoff, Niklas and Kocetkov, Denis and Mou, Chenghao and Marone, Marc and Akiki, Christopher and Li, Jia and Chim, Jenny and others},
  journal={arXiv preprint arXiv:2305.06161},
  year={2023}
}

@article{lozhkov2024starcoder,
  title={Starcoder 2 and the stack v2: The next generation},
  author={Lozhkov, Anton and Li, Raymond and Allal, Loubna Ben and Cassano, Federico and Lamy-Poirier, Joel and Tazi, Nouamane and Tang, Ao and Pykhtar, Dmytro and Liu, Jiawei and Wei, Yuxiang and others},
  journal={arXiv preprint arXiv:2402.19173},
  year={2024}
}

@article{guo2024deepseek,
  title={DeepSeek-Coder: When the Large Language Model Meets Programming--The Rise of Code Intelligence},
  author={Guo, Daya and Zhu, Qihao and Yang, Dejian and Xie, Zhenda and Dong, Kai and Zhang, Wentao and Chen, Guanting and Bi, Xiao and Wu, Yu and Li, YK and others},
  journal={arXiv preprint arXiv:2401.14196},
  year={2024}
}

@article{puri2021codenet,
  title={Codenet: A large-scale ai for code dataset for learning a diversity of coding tasks},
  author={Puri, Ruchir and Kung, David S and Janssen, Geert and Zhang, Wei and Domeniconi, Giacomo and Zolotov, Vladimir and Dolby, Julian and Chen, Jie and Choudhury, Mihir and Decker, Lindsey and others},
  journal={arXiv preprint arXiv:2105.12655},
  year={2021}
}

@inproceedings{gandhi2024translation,
  title={Translation of low-resource COBOL to logically correct and readable Java leveraging high-resource Java refinement},
  author={Gandhi, Shubham and Patwardhan, Manasi and Khatri, Jyotsana and Vig, Lovekesh and Medicherla, Raveendra Kumar},
  booktitle={Proceedings of the 1st International Workshop on Large Language Models for Code},
  pages={46--53},
  year={2024}
}

@article{hui2024qwen2,
  title={Qwen2. 5-coder technical report},
  author={Hui, Binyuan and Yang, Jian and Cui, Zeyu and Yang, Jiaxi and Liu, Dayiheng and Zhang, Lei and Liu, Tianyu and Zhang, Jiajun and Yu, Bowen and Lu, Keming and others},
  journal={arXiv preprint arXiv:2409.12186},
  year={2024}
}

@article{cassano2023multipl,
  title={Multipl-e: A scalable and polyglot approach to benchmarking neural code generation},
  author={Cassano, Federico and Gouwar, John and Nguyen, Daniel and Nguyen, Sydney and Phipps-Costin, Luna and Pinckney, Donald and Yee, Ming-Ho and Zi, Yangtian and Anderson, Carolyn Jane and Feldman, Molly Q and others},
  journal={IEEE Transactions on Software Engineering},
  volume={49},
  number={7},
  pages={3675--3691},
  year={2023},
  publisher={IEEE}
}

@article{lankford2023adaptmllm,
  title={adaptmllm: Fine-tuning multilingual language models on low-resource languages with integrated llm playgrounds},
  author={Lankford, S{\'e}amus and Afli, Haithem and Way, Andy},
  journal={Information},
  volume={14},
  number={12},
  pages={638},
  year={2023},
  publisher={MDPI}
}

@article{song2025llm,
  title={Is LLM the Silver Bullet to Low-Resource Languages Machine Translation?},
  author={Song, Yewei and Li, Lujun and Lothritz, Cedric and Ezzini, Saad and Sleem, Lama and Gentile, Niccolo and State, Radu and Bissyand{\'e}, Tegawend{\'e} F and Klein, Jacques},
  journal={arXiv preprint arXiv:2503.24102},
  year={2025}
}

@article{luo2025unlocking,
  title={Unlocking LLM Repair Capabilities in Low-Resource Programming Languages Through Cross-Language Translation and Multi-Agent Refinement},
  author={Luo, Wenqiang and Keung, Jacky Wai and Yang, Boyang and Klein, Jacques and Bissyande, Tegawende F and Tian, Haoye and Le, Bach},
  journal={arXiv preprint arXiv:2503.22512},
  year={2025}
}

@article{sontakke2023knowledge,
  title={Knowledge Transfer for Pseudo-code Generation from Low Resource Programming Language},
  author={Sontakke, Ankita and Kalra, Kanika and Patwardhan, Manasi and Vig, Lovekesh and Medicherla, Raveendra Kumar and Naik, Ravindra and Pradhan, Shrishti},
  journal={arXiv preprint arXiv:2303.09062},
  year={2023}
}

@article{joel2024survey,
  title={A survey on llm-based code generation for low-resource and domain-specific programming languages},
  author={Joel, Sathvik and Wu, Jie and Fard, Fatemeh},
  journal={ACM Transactions on Software Engineering and Methodology},
  year={2024},
  publisher={ACM New York, NY}
}

@article{alyami2025domain,
  title={Domain-Adaptive Pre-Training for Arabic Aspect-Based Sentiment Analysis: A Comparative Study of Domain Adaptation and Fine-Tuning Strategies},
  author={Alyami, Salha and Jamal, Amani and Alhothali, Areej},
  journal={arXiv preprint arXiv:2509.16788},
  year={2025}
}

@article{yan2025adaft,
  title={AdaFT: An efficient domain-adaptive fine-tuning framework for sentiment analysis in chinese financial texts},
  author={Yan, Guofeng and Peng, Kuashuai and Wang, Yongfeng and Tan, Hengliang and Du, Jiao and Wu, Heng},
  journal={Applied Intelligence},
  volume={55},
  number={7},
  pages={701},
  year={2025},
  publisher={Springer}
}

@misc{noauthor_bloop_nodate,
	title = {bloop {\textbar} {Evaluating} {LLMs} on {COBOL}},
	url = {https://bloop.ai/blog/evaluating-llms-on-cobol},
	language = {en},
	urldate = {2025-02-16},
author={BloopAI},
year={2024}
}

@inproceedings{zheng2023codegeex,
  title={Codegeex: A pre-trained model for code generation with multilingual benchmarking on humaneval-x},
  author={Zheng, Qinkai and Xia, Xiao and Zou, Xu and Dong, Yuxiao and Wang, Shan and Xue, Yufei and Shen, Lei and Wang, Zihan and Wang, Andi and Li, Yang and others},
  booktitle={Proceedings of the 29th ACM SIGKDD Conference on Knowledge Discovery and Data Mining},
  pages={5673--5684},
  year={2023}
}

@article{zhuo2024bigcodebench,
  title={Bigcodebench: Benchmarking code generation with diverse function calls and complex instructions},
  author={Zhuo, Terry Yue and Vu, Minh Chien and Chim, Jenny and Hu, Han and Yu, Wenhao and Widyasari, Ratnadira and Yusuf, Imam Nur Bani and Zhan, Haolan and He, Junda and Paul, Indraneil and others},
  journal={arXiv preprint arXiv:2406.15877},
  year={2024}
}

@misc{cobolcodebench,
	title = {},
	url = {https://huggingface.co/datasets/harshini-kumar/CobolCodeBench},
	language = {en},
	urldate = {2025-02-16},
author={Harshini Kumar},
year={2025}
}

@article{team2024qwen2,
  title={Qwen2 technical report},
  author={Team, Qwen and others},
  journal={arXiv preprint arXiv:2407.10671},
  volume={2},
  number={3},
  year={2024}
}

@article{austin2021program,
  title={Program synthesis with large language models},
  author={Austin, Jacob and Odena, Augustus and Nye, Maxwell and Bosma, Maarten and Michalewski, Henryk and Dohan, David and Jiang, Ellen and Cai, Carrie and Terry, Michael and Le, Quoc and others},
  journal={arXiv preprint arXiv:2108.07732},
  year={2021}
}

@article{chen2021codex,
  title={Evaluating Large Language Models Trained on Code},
  author={Mark Chen and Jerry Tworek and Heewoo Jun and Qiming Yuan and Henrique Ponde de Oliveira Pinto and Jared Kaplan and Harri Edwards and Yuri Burda and Nicholas Joseph and Greg Brockman and Alex Ray and Raul Puri and Gretchen Krueger and Michael Petrov and Heidy Khlaaf and Girish Sastry and Pamela Mishkin and Brooke Chan and Scott Gray and Nick Ryder and Mikhail Pavlov and Alethea Power and Lukasz Kaiser and Mohammad Bavarian and Clemens Winter and Philippe Tillet and Felipe Petroski Such and Dave Cummings and Matthias Plappert and Fotios Chantzis and Elizabeth Barnes and Ariel Herbert-Voss and William Hebgen Guss and Alex Nichol and Alex Paino and Nikolas Tezak and Jie Tang and Igor Babuschkin and Suchir Balaji and Shantanu Jain and William Saunders and Christopher Hesse and Andrew N. Carr and Jan Leike and Josh Achiam and Vedant Misra and Evan Morikawa and Alec Radford and Matthew Knight and Miles Brundage and Mira Murati and Katie Mayer and Peter Welinder and Bob McGrew and Dario Amodei and Sam McCandlish and Ilya Sutskever and Wojciech Zaremba},
  year={2021},
  eprint={2107.03374},
  archivePrefix={arXiv},
  journal={arXiv preprint arXiv:2107.03374},
  primaryClass={cs.LG}
}

@misc{taulli_cobol_nodate,
	title = {{COBOL} {Language}: {Call} {It} {A} {Comeback}?},
	shorttitle = {{COBOL} {Language}},
	url = {https://www.forbes.com/sites/tomtaulli/2020/07/13/cobol-language-call-it-a-comeback/},
	language = {en},
	urldate = {2025-02-16},
	journal = {Forbes},
	author = {Taulli, Tom},
        year={2020}
}

@article{yan2023codetransocean,
  title={Codetransocean: A comprehensive multilingual benchmark for code translation},
  author={Yan, Weixiang and Tian, Yuchen and Li, Yunzhe and Chen, Qian and Wang, Wen},
  journal={arXiv preprint arXiv:2310.04951},
  year={2023}
}

@article{liu2023your,
  title={Is your code generated by chatgpt really correct? rigorous evaluation of large language models for code generation},
  author={Liu, Jiawei and Xia, Chunqiu Steven and Wang, Yuyao and Zhang, Lingming},
  journal={Advances in Neural Information Processing Systems},
  volume={36},
  pages={21558--21572},
  year={2023}
}

@article{froimovich2025quality,
  title={Quality Evaluation of COBOL to Java Code Transformation},
  author={Froimovich, Shmulik and Gal, Raviv and Ibraheem, Wesam and Ziv, Avi},
  journal={arXiv preprint arXiv:2507.23356},
  year={2025}
}

@inproceedings{kumar2024automated,
  title={Automated validation of cobol to java transformation},
  author={Kumar, Atul and Saha, Diptikalyan and Yasue, Toshiaki and Ono, Kohichi and Krishnan, Saravanan and Hans, Sandeep and Satoh, Fumiko and Mitchell, Gerald and Kumar, Sachin},
  booktitle={Proceedings of the 39th IEEE/ACM International Conference on Automated Software Engineering},
  pages={2415--2418},
  year={2024}
}

@inproceedings{lee2022opencbs,
  title={OpenCBS: An Open-Source COBOL Defects Benchmark Suite},
  author={Lee, Dylan and Henley, Austin Z and Hinshaw, Bill and Pandita, Rahul},
  booktitle={2022 IEEE International Conference on Software Maintenance and Evolution (ICSME)},
  pages={246--256},
  year={2022},
  organization={IEEE}
}

@misc{gpt4omini,
        author={OpenAI},
	title = {GPT-4o mini: advancing cost-efficient intelligence},
	url = {https://openai.com/index/gpt-4o-mini-advancing-cost-efficient-intelligence/},
	language = {en},
	urldate = {2025-02-16},
    year={2024}
}

@misc{claude,
        author={Anthropic},
	title = {Claude Sonnet 4.5},
	url = {https://www.anthropic.com/claude/sonnet},
	language = {en},
	urldate = {2025-02-16},
    year={2025}
}

@misc{gptoss,
        author={OpenAI},
	title = {Introducing gpt-oss},
	url = {https://openai.com/index/introducing-gpt-oss/},
	language = {en},
	urldate = {2025-10-16},
    year={2025}
}

@article{yu2023large,
  title={Large language model as attributed training data generator: A tale of diversity and bias},
  author={Yu, Yue and Zhuang, Yuchen and Zhang, Jieyu and Meng, Yu and Ratner, Alexander J and Krishna, Ranjay and Shen, Jiaming and Zhang, Chao},
  journal={Advances in neural information processing systems},
  volume={36},
  pages={55734--55784},
  year={2023}
}

@article{wei2023magicoder,
  title={Magicoder: Empowering code generation with oss-instruct},
  author={Wei, Yuxiang and Wang, Zhe and Liu, Jiawei and Ding, Yifeng and Zhang, Lingming},
  journal={arXiv preprint arXiv:2312.02120},
  year={2023}
}

@article{lei2025enhancing,
  title={Enhancing COBOL Code Explanations: A Multi-Agents Approach Using Large Language Models},
  author={Lei, Fangjian and Liu, Jiawen and Noei, Shayan and Zou, Ying and Truong, Derek and Alexander, William},
  journal={arXiv preprint arXiv:2507.02182},
  year={2025}
}

@article{borges2018s,
  title={What’s in a github star? understanding repository starring practices in a social coding platform},
  author={Borges, Hudson and Valente, Marco Tulio},
  journal={Journal of Systems and Software},
  volume={146},
  pages={112--129},
  year={2018},
  publisher={Elsevier}
}

@article{ali2023x,
  title={X-cobol: A dataset of cobol repositories},
  author={Ali, Mir Sameed and Manjunath, Nikhil and Chimalakonda, Sridhar},
  journal={arXiv preprint arXiv:2306.04892},
  year={2023}
}

\appendix
\section*{Appendix}
\section{Prompts Used in the Data Construction Pipeline}
\label{appendix:prompt}

\subsection{Compiler-Guided Repair Prompt}
\begin{tcolorbox}
You are an experienced COBOL software engineer with deep knowledge of COBOL syntax, structure, and best practices. Your task is to perform debugging on a given COBOL program with the compilation errors.

Below is the original COBOL code followed by the compiler’s error log. Your job is to revise the code to resolve all compilation errors, ensuring that the corrected program is not only syntactically valid but also logically sound.

Please carefully analyze the error messages and update the code accordingly. Prioritize clarity, maintainability, and adherence to COBOL’s structural rules.

Input:

COBOL Code:

\textit{[COBOL code]}

Compiler Error Log:

\textit{[Error log]}

\end{tcolorbox}

\subsection{LLM-based Pair Scoring Prompt}

\begin{tcolorbox}
You are an expert software engineer with deep knowledge of both COBOL and Java.
Your task is to evaluate whether the following two programs are semantically equivalent.

Program 1 (COBOL):

\textit{[COBOL Code]}

Program 2 (Java):

\textit{[Java Code]}

Evaluate their similarity based on the following criteria:

1. Do both programs implement the same functionality?

2. Do they produce the same outputs for the same inputs?

3. Are there any logical differences, missing steps, or incorrect translations?

4. Ignore differences in syntax, formatting, or variable naming.

Scoring:

- 1.0 = Fully equivalent (same logic and behavior)

- 0.7-0.9 = Minor differences but mostly equivalent

- 0.4-0.6 = Partial similarity (some logic mismatch)

- 0.0–0.3 = Not equivalent

Output your answer in the following format:

Score: [number between 0 and 1.0]

Explanation: [brief explanation of reasoning]
\end{tcolorbox}

\subsection{Back Translation Prompt}

\begin{tcolorbox}
    You are a senior engineer specializing in COBOL and Java. 
Translate the following COBOL snippet into Java.

CRITICAL CONSTRAINTS (follow strictly)

1. Minimalism first: output ONLY the Java code required to reflect what appears in the COBOL snippet.

2. Do NOT add any “helpful” extras: no getters/setters, no beans, no padding utilities, no data validation, no additional methods, no comments explaining mapping, no test plan, no assumptions section, no package/imports unless strictly required for compilation.

3. Do NOT invent structure or frameworks. If COBOL defines data items but they are never used by PROCEDURE DIVISION logic, you must NOT create Java classes for them. (You may ignore unused data definitions entirely.)

4. If the COBOL PROCEDURE DIVISION only prints messages and stops, the Java output should only contain a single class with main() that prints the same messages and returns.

5. Preserve literals and observable behavior exactly:

   - DISPLAY to System.out.print/println (choose print vs println to best match; default to println unless COBOL shows no newline requirement).
   
   - STOP RUN to return from main (no System.exit unless COBOL implies abnormal termination).
   
6. Keep formatting simple and close to typical “plain Java” style:

7. If the COBOL is missing required info for a valid Java identifier or class name, use the closest safe name and do not add explanations.

OUTPUT FORMAT
    
Return ONE Java code block only, no additional text.

Here is the given COBOL code:

\textit{[COBOL code]}
\end{tcolorbox}

\subsection{Instruction Generation Prompt}
\begin{tcolorbox}
    You are exceptionally skilled at crafting high-quality COBOL programming problems and offering precise solutions.
    
Please gain inspiration from the following code snippet to create a high-quality programming problem.

If the given code is not COBOL, return None

Otherwise, present your output in two distinct sections: [Problem Description] and [Solution].

Code snippet for inspiration: 

 \textit{[COBOL code]} 

Guidelines for each section:

1. [Problem Description]: This should be completely self-contained, providing all the contextual information one needs to understand and solve the problem. Assume common programming knowledge, but ensure that any specific context, variables, or code snippets pertinent to this problem are explicitly included.

2. [Solution]: Offer a comprehensive, correct solution that accurately addresses the [Problem Description] you provided.

\end{tcolorbox}

\subsection{Candidate Selection Prompt}
\begin{tcolorbox}
    You are an expert in designing high-quality COBOL programming problems and providing accurate solutions.
    
You will be given four problem descriptions, each corresponding to the same COBOL code snippet.
Your task is to carefully evaluate the given options and select the most suitable problem description based on its clarity, relevance, and alignment with the provided code.

Instructions:

1. Analyze all three options thoroughly.

2. Select the option that best matches and explains the code.

3. At the end of your response, indicate your choice in the following format: [Best option: X], where X is the number of the selected option.

Code snippet for reference: 

\textit{ [COBOL code]}  

Option 1: [\textit{{Description 1}}]

Option 2: [\textit{{Description 2}}]

Option 3: [\textit{{Description 3}}]

Option 4: [\textit{{Description 4}}]
\end{tcolorbox}

\section{Similarity Score Distributions}

\subsection{LLM-based Pair Scoring Distribution}
\begin{figure*}[ht]
\centering
\includegraphics[width=\textwidth]{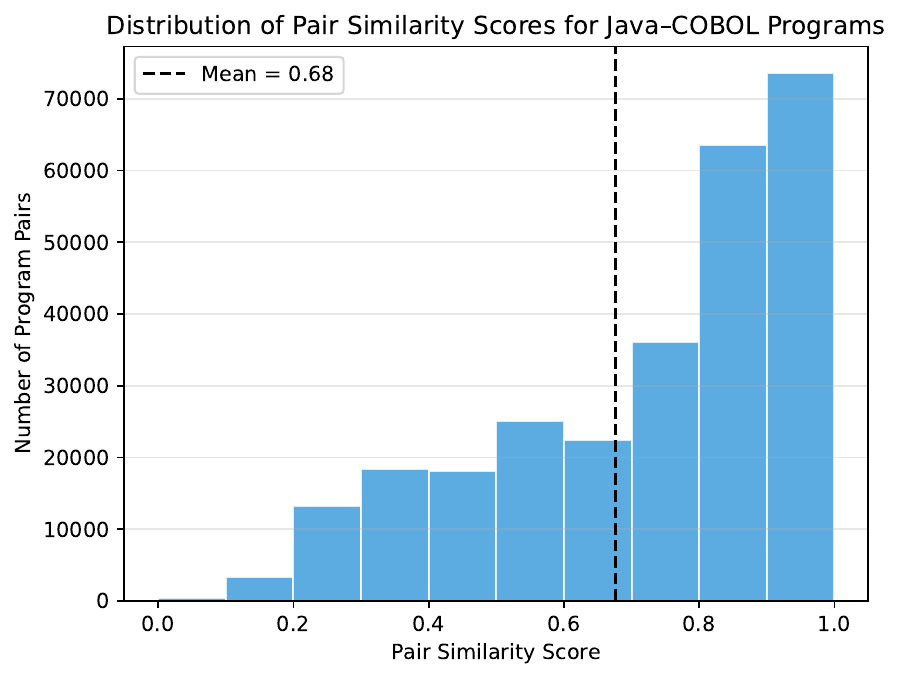}
\caption{Distribution of Pair Similarity Scores in LLM-based Pair Scoring (Java–COBOL)}
\label{fig:pair_scoring}
\end{figure*}

\subsection{CodeBERTScore Distribution (with vs without normalization)}
\begin{figure*}[ht]
\centering
\includegraphics[width=\textwidth]{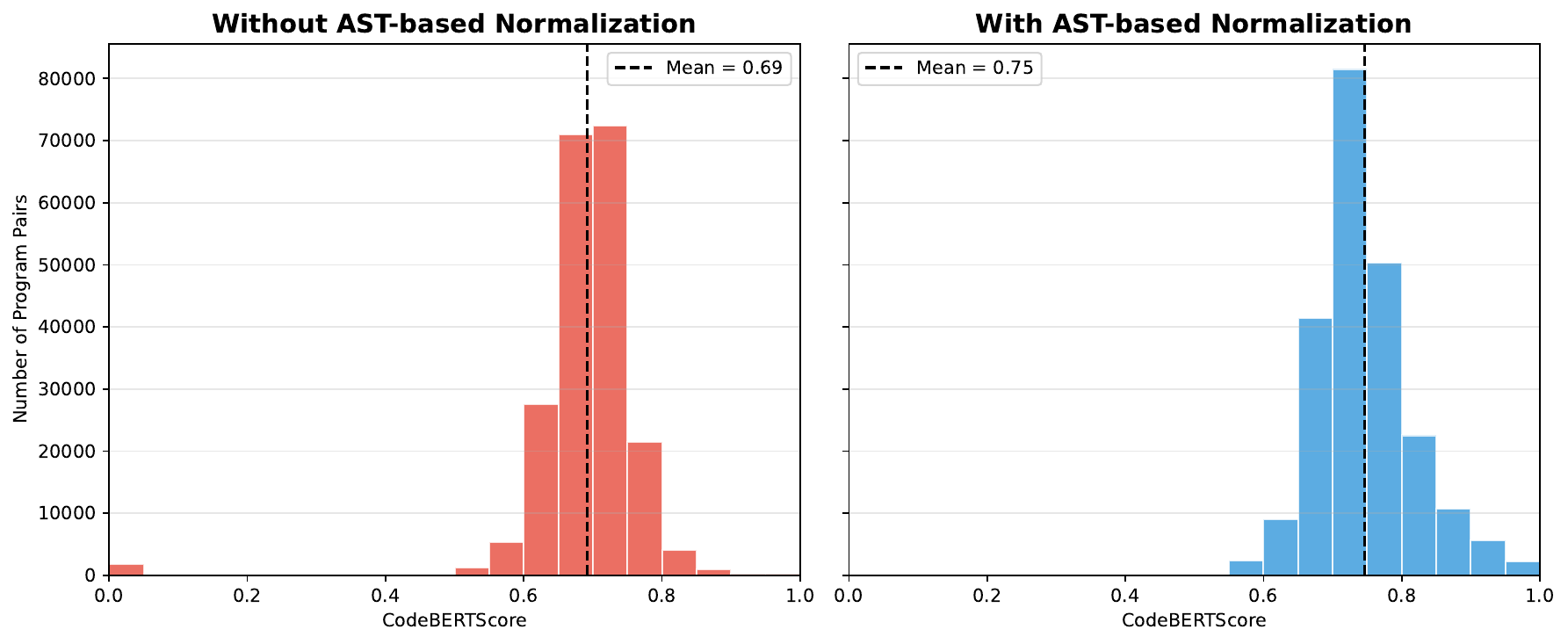}
\caption{Distribution of CodeBERTScore With and Without AST-based normalization.}
\label{fig:codebert}
\end{figure*}
\end{document}